\documentclass[english,12pt]{article}
\usepackage{array}
\usepackage{graphicx}
\usepackage{amssymb}
\usepackage{amsmath}
\usepackage{multirow}
\usepackage{prettyref}
\usepackage{babel}
\usepackage{units}
\usepackage[latin1]{inputenc}
\usepackage{amsfonts}
\usepackage{amssymb}
\usepackage{babel}
\usepackage{color}
\usepackage{cite}
\def\@fmsl@sh#1#2#3{\m@th\ooalign{$\hfil#1\mkern#2/\hfil$\crcr$#1#3$}}
 \def\eq#1\en{\begin{equation}#1\end{equation}}
\def\s[#1,#2]{[#1\stackrel{\star}{,}#2]}
\def\sx[#1,#2]{[#1\stackrel{\star_{x}}{,}#2]}

\newcommand{\nc}{\newcommand}
\nc{\beq}{\begin{equation}}
\nc{\eeq}{\end{equation}}
\nc{\beqa}{\begin{eqnarray}}
\nc{\eeqa}{\end{eqnarray}}

\def\bc{\begin{center}}
\def\ec{\end{center}}

\def\gsim{\mathrel{\mathpalette\atversim>}}

\def\bc{\begin{center}}
\def\ec{\end{center}}

\def\gsim{\mathrel{\rlap{\lower4pt\hbox{\hskip1pt$\sim$}}\raise1pt\hbox{$>$}}}       
\def\gsim{\mathrel{\rlap{\lower4pt\hbox{\hskip1pt$\sim$}}\raise1pt\hbox{$>$}}}       
\begin{document}
\makeatletter
\def\fmslash{\@ifnextchar[{\fmsl@sh}{\fmsl@sh[0mu]}}
\def\fmsl@sh[#1]#2{%
  \mathchoice
    {\@fmsl@sh\displaystyle{#1}{#2}}%
    {\@fmsl@sh\textstyle{#1}{#2}}%
    {\@fmsl@sh\scriptstyle{#1}{#2}}%
    {\@fmsl@sh\scriptscriptstyle{#1}{#2}}}
\def\@fmsl@sh#1#2#3{\m@th\ooalign{$\hfil#1\mkern#2/\hfil$\crcr$#1#3$}}
\makeatother
%
\thispagestyle{empty}
\begin{titlepage}
\boldmath
\begin{center}
  \Large {\bf What is the final state of a black hole merger?}
    \end{center}
\unboldmath
\vspace{0.2cm}
\begin{center}
{ {\large Xavier Calmet}\footnote{x.calmet@sussex.ac.uk}$^{a,b}$
{\large and}
{\large Roberto Casadio}\footnote{casadio@bo.infn.it}$^{c,d}$  }
 \end{center}
\begin{center}
{\sl  $^a$Department of Physics and Astronomy, 
University of Sussex, Falmer, Brighton, BN1 9QH, United Kingdom 
}
\\
$^b${\sl PRISMA Cluster of Excellence and Mainz Institute for Theoretical Physics, Johannes Gutenberg University,
55099 Mainz, Germany }
\\
{\sl  $^c$Dipartimento di Fisica e Astronomia, Universit\`a di Bologna, via Irnerio~46, I-40126 Bologna, Italy}
\\
{\sl $^d$I.N.F.N., Sezione di Bologna, via B.~Pichat~6/2, I-40127 Bologna, Italy}
\end{center}
\vspace{5cm}
\begin{abstract}
\noindent
In this short paper we discuss the possibility of testing the nature of astrophysical black holes using the recently observed
black hole mergers.
We investigate the possibility that a secondary black hole is created in the merger of two astrophysical black holes and
discuss potential astrophysical signatures.
We point out that black hole mergers are a possible astrophysical mechanism for the creation of quantum black holes
with masses close to the Planck mass. 
\end{abstract}  
\end{titlepage}
%
%
%
%
\newpage
\section{Introduction}
The recent observations of gravitational waves by the LIGO and Virgo Scientific
collaborations~\cite{Abbott:2016blz,Abbott:2016nmj,Abbott:2017vtc}
provide a new window to study our universe and black holes in particular.
Black hole physics is fascinating as it requires to understand gravity in the strong field regime.
The study of black holes thus offers new opportunities to test gravity in that regime which has not been probed much so far.
In particular, the question of black hole formation is an extremely interesting but complicated topic.
While in vacuum exact static solutions to Einstein's equations such as the Schwarzschild solution or the Kerr-Schild metric can
be found easily, real astrophysical black holes are not in vacuum as they are surrounded by matter.
Furthermore, a real black hole is clearly a dynamical object that is the result of a dynamical process,
namely the gravitational collapse of a region of space-time.
One therefore has to rely on other criteria, such as the Hoop conjecture~\cite{hoop} or the mathematically more
rigorous argument based on the formation of a closed trapped surface, in order to claim that a region of space-time
contains an apparent horizon and is thus a black hole in formation (see, e.g.,~Ref.~\cite{Ashtekar:2004cn} for a
comprehensive review of dynamical horizons). 
\par
This raises an even more basic question, namely that of understanding what astrophysical black holes truly are.
Clearly they cannot be strictly eternal black holes.
They will eventually become eternal black holes given an infinite amount of time to evolve, but at this stage of the universe,
astrophysical black holes are not eternal black holes as they are not in vacuum.
But how close are they from being eternal?~\footnote{A simple back of the envelop calculation using textbook general relativity
and a simple model for the collapse of a star shows that within a very short period of time the outer layer of the collapsing
star will have reached a distance less than the Planck size from the horizon of the black hole.
However, it takes an infinite amount of time for all of the collapsing matter to actually cross the horizon, from the point of view
of a distant observer.}
Interestingly, there is a simple observational signature which can help us to answer this question.
Hawking has shown~\cite{Hawking:1971vc,Misner:1974qy} that the area of a black hole can never decrease classically.
This derivation applies to classical general relativity, and does not account for the Hawking radiation~\cite{Hawking:1974sw}.
In fact, the derivation of Hawking's classical result is based on the assumption that black holes are eternal black holes and,
if astrophysical black holes are not yet eternal black holes, small departures from the area conservation law of black holes
are possible.
The importance of distinguishing between physical and eternal black holes has recently been stressed in the context of the
calculation of quantum corrections to black holes~\cite{Calmet:2017qqa}.
We will here propose an observational test of the nature of black holes using the recent discovery of gravitational waves.
\section{Merging and apparent horizons}
The gravitational wave signals GW150914, GW151226 and  GW170104 detected by the LIGO detectors are interpreted
as originating from a pair of merging black holes.
Calculating the gravitational-wave signal for such merging black holes is done using numerical relativity simulations,
because analytic approximations fail near the time of the merger~\footnote{Note that some analytical solutions are known
in the case of a large mass asymmetry between the two merging black holes~\cite{Hamerly:2010cr,Emparan:2016ylg}.},
i.e.~when the black holes are very close to each other.
These simulations are designed to show that an apparent horizon forms when the two black holes merge.
\par
Less relevant for astrophysical black holes, but just as fascinating from a theoretical point of view,
it is possible to prove analytically that black holes can form in a high energy collision of two boosted
Schwarzschild metrics~\cite{Penrose,DEath:1992plq,Eardley:2002re}.
In that case it is possible to show analytically the existence of an apparent horizon even for a non-zero impact
parameter (essentially in agreement with the hoop conjecture~\cite{hoop}).
\par
Proving the existence of an apparent horizon is the key element of all studies of the dynamics of black hole formation.
It is however important to keep in mind that the existence of an apparent horizon does not imply that only one black hole
is formed when two trapped surfaces collide or merge, rather several black holes could be formed in such processes.
The collision of two black holes could be akin to the collision of a large comet or another large celestial body with a
planet thereby sometime creating a satellite to the planet.
The usual assumption with black hole collisions or mergers is however that only one large black hole is formed
(see, e.g.,~the comments in Ref.~\cite{DEath:1992plq,Eardley:2002re}), but this does not esclude the temporary
formation of more apparent horizons which will be hidden inside the final event horizon.
Furthermore, as explained previously, the uniqueness of the final state is expected in case of a merger between
two eternal black holes because of the area law for black holes.
These assumptions have also been made by the LIGO collaboration and they are perfectly compatible with
numerical simulations.
The aim of this work is to argue that we have for the first time observational data to verify these key assumptions
of black hole formation.
\par
Independently of our argument based on the nature of real astrophysical black holes in general relativity,
models of black holes beyond general relativity could also lead to a violation of the area law of black holes.
This possibility becomes generically more likely already in the semiclassical approximation in which space-time
can still be viewed as classical geometry but matter is described by quantum physics.
In particular, the energy conditions required by the area theorems might not be satisfied by all of the matter in
the system.
Indeed, we have already recalled that is the case for the Hawking radiation, which one expects appears in the
proximity of any apparent horizons~\cite{Hajicek:1986hn}.
As a consequence, apparent horizons can also decrease in size and disappear after they have formed.
Even if only one large black hole is left after the merging, one cannot esclude the temporary formation
of more apparent horizons, which disappear before the final event horizon develops.
In fact, that the merging of two black holes is more complicated than usually assumed fits well
with recent speculations (see e.g.~Ref.~\cite{Almheiri:2012rt}) on the possibility that the horizon of black holes
might be a less trivial region of space-time than the one general relativity seems to predict.
For instance, a firewall might be present which acts as a physical surface in contrast to the usual horizon of
a black hole.
In that sense the horizon might be more like the surface of a planet than empty space as forecasted by
general relativity. 
\par
The GW150914 observation~\cite{Abbott:2016blz} is interpreted as the merger of two black holes of masses
$36^{+5}_{-4}\, M_{\odot}$ and $29^{+4}_{-4}\, M_{\odot}$ where $M_{\odot}$ stands for the solar mass.
The final black hole has a mass of $62^{+4}_{-4}\,M_{\odot}$ and the energy radiated in form of gravitational waves
is estimated to be of the order of $3^{+0.5}_{-0.5}\, M_{\odot} c^2$.
Clearly these numbers beautifully verify the assumption that only one black hole is formed in the merger.
However, they are also compatible, in an extreme case, with a final state of two black holes, one with a heavy mass
of the order of $56\, M_{\odot}$ and a lighter one of the order of $6\, M_{\odot}$.
There is a wide range for the mass of a potential secondary black hole.
Its mass could be from the lightest possible one, i.e.~a Planckian quantum black hole of mass of the order of the Planck
mass $M_P\simeq 2.17\times 10^{-8}\, \mbox{kg}=1\times 10^{-36}\,M_{\odot}$~\footnote{Note that the Planck scale is a
dynamical quantity which depends on the number of fields in the theory, the true Planck mass could be well below the
value obtained by dimensional analysis $2.17 \times 10^{-8} \mbox{kg}$~\cite{Calmet:2014gya}.}
to the order of $6\, M_{\odot}$ mentioned above.
The secondary black hole will be a classical black hole if its mass is below $6\, M_{\odot}$ but much larger than $M_P$.
It will be a semiclassical black hole if its mass is of the order of $20\,M_P$ and a quantum black hole below that.
Note that the theoretical existence of quantum black holes is well motivated by results obtained using effective field theory
methods applied to quantum gravity~\cite{Calmet:2014gya,Calmet:2015pea,Calmet:2017omb,Calmet:2015fua,Calmet:2014dea}.
\par
Mergers of black holes are thus potentially a mechanism for the formation of black holes with masses below the well-known
2-3$\,M_{\odot}$ threshold for astrophysical black holes but also for Planckian black holes.
The fate of the secondary black hole could be very different depending on its mass.
However, while the true nature of astrophysical black holes is questionable, classical general relativity tells us that since
the two merging black holes are gravitationally bound, if there is a secondary black hole produced, whatever its mass might be
it will be gravitationally bound as well to the main large black hole.
This implies that in all likelihood, the first merger could be followed by a secondary merger within a short amount of time,
which would in turn lead to a second gravitational wave detection this time corresponding to secondary black hole merging
with the large one.
This will be the case if the secondary black hole can be treated as a classical black hole.
Whatever the mass of the secondary black hole, we expect it to be captured by the primary one faster than it would have
time to potentially decay via Hawking radiation.
\par
For classical black holes, we can use the famous result by Hawking to calculate their temperature~\cite{Hawking:1974sw}.
In SI~units, a Schwarzschild black hole has a temperature given by
\begin{eqnarray}
T
=
\frac{\hbar \,c^3}{8\, \pi\, G\, M\, k_B}
=
6.169 \times 10^{-8}\, \mbox{K} \times \frac{M_{\odot}}{M}
\ .
\label{Th} 
\end{eqnarray}
where $\hbar$ is the reduced Planck constant, $c$ is the speed of light,  $G$ is Newton's constant, $k_B$ is
Boltzmann's constant, $M$ the mass of the black hole and the temperature is calculated in Kelvin (K).
The current temperature of the universe is of $2.7\,$K.
This implies that only black holes with a mass below $2.28 \times 10^{-8}\,M_{\odot}$ will radiate, the heavier one
will remain stable.
In both cases however we expect that the secondary black hole will merge into the primary one faster than it would
need time to decay.
\par
The ultimate fate of secondary black holes with masses below $2.28 \times 10^{-8}\,M_{\odot}$ depends on whether
the lightest black hole is a stable remnant or not.
All black holes with a mass below $2.28 \times 10^{-8}\,M_{\odot}$ will lose mass via Hawking radiation and ultimately
end up as a Planckian black hole.
Note, however, that it would take $2.4 \times 10^{44}\,$years for a black hole of mass $2.28 \times 10^{-8}\,M_{\odot}$
to radiate down to a Planckian black hole according to the standard Hawking formula~\eqref{Th}
for which the black hole life-time is proportional to $M^3$.
Planckian black holes belong to the realm of quantum gravity and it is thus difficult to make model independent
predictions for their physics.
In particular, it is not clear whether the lightest Planckian black holes could be a remnant, i.e.~a stable object with a mass
close to the Planck mass, or whether they will decay into particles of the standard model.
Remnants have been heavily criticized, probably mainly for aesthetical reasons, but they remain a valid option~\cite{Calmet:2014uaa}
from an experimental point of view.
However, recent progress using effective field theory techniques indicates that the lightest states in quantum gravity,
which can be interpreted as black hole precursors~\cite{Calmet:2014gya,Calmet:2015pea}, should be very short-lived
objects~\cite{Calmet:2014gya}.
\section{Conclusions and outlook}
So far we have argued about the complexity of the merging of two black holes, but we would like to conclude with
some more specific (albeit speculative) observational prediction. 
If, as expected, the lightest black hole is a very short-lived object, akin to a resonance with a decay width $\Gamma$
of the order of Planck mass $M_P$ and a lifetime of the order of $1/\Gamma$~\cite{Calmet:2014gya,Calmet:2015pea},
there is a very clear signal for the decay of the final state secondary black hole.
This would happen before the secondary black hole can be recaptured by the primary one.
The secondary Planckian black hole is a non-thermal object and it will thus typically decay only in a couple of particles.
While the Planckian black hole will decay into fermions and electroweak gauge bosons or gravitons, the most interesting
decay mode is into photons as these energetic particles should be observable.
We can estimate the branching ratio for the black hole decaying into a pair of photons by simply counting the number of
possible final states, and we find it is about 0.5$\%$. 
\par
If the final state of the merger consists of a large black hole and a Planckian object, the gravitational wave signal from the
merger should be accompanied by a flash of light of very high energy, each photon carrying $1.8 \times 10^{16}\,$GeV.
These bursts would be extremely short, and last only $9\times 10^{-44}\,$s (i.e.~the decay time of the black hole) and
thus probably very difficult to observe.
Other high energetic pairs of fermion anti-fermion (electrons, muons, taus, quarks) with similar energies to the photons
discussed above would also be produced and could be a source of  high energetic cosmic rays.
\par
Our prediction of a burst of energy in the electromagnetic spectrum is not unique.
For instance, it should also occur during the black hole to white hole transition~\cite{Barrau:2014yka}.
However, in that case, the burst of gamma rays would correspond to the complete disappearance of the whole
system, whereas for the effect we consider here a large black hole would remain behind.

While we acknowledge that the physics of quantum black holes is inherently speculative as we do not yet have a fully satisfactory UV complete theory of quantum gravity, our current knowledge of black holes leaves open the possibility that astrophysical mergers of black holes could be more subtle than usually assumed. We have proposed potential observational tests of such a possibility.
\subsection*{Acknowledgments}
The work of XC is supported in part  by the Science and Technology Facilities Council (grant number ST/P000819/1).
XC is very grateful to MITP for their generous hospitality during the academic year 2017/2018.
The work of RC is partially supported by the INFN grant FLAG.
%

\bigskip{}

\end{document}